\documentclass[useAMS,usenatbib,usegraphicx,usexfrac]{mn2e}

\def\Msun{\hbox{$\rmn{M}_{\sun}~$}}
\def\Mbh{\hbox{$M_{\rmn{BH}}$}}

\title[UV and X-ray variability of NGC 4051]
  {Ultraviolet and X-ray variability of NGC 4051 over 45 days with \textit{XMM-Newton} and \textit{Swift}}
\author[W. N. Alston et al.]
  {W. N. Alston$^{1}$\thanks{E-mail: wna3@leicester.ac.uk}, S. Vaughan$^{1}$ and P. Uttley$^{2}$ \\
  $^{1}$X-Ray and Observational Astronomy Group, University of Leicester, Leicester, LE1 7RH. UK\\
  $^{2}$Astronomical Institute “Anton Pannekoek”, University of Amsterdam, Postbus 94249, 1090 GE Amsterdam, The Netherlands}

\date{Accepted 2012 October 30. Recieved 2012 October 24; in original form 2012 October 9}

\pagerange{\pageref{firstpage}--\pageref{lastpage}} \pubyear{2012}

\def\LaTeX{L\kern-.36em\raise.3ex\hbox{a}\kern-.15em
    T\kern-.1667em\lower.7ex\hbox{E}\kern-.125emX}

\begin{document}
\label{firstpage}
\maketitle

\begin{abstract}
We analyse 15 \textit{XMM-Newton} observations of the Seyfert galaxy NGC 4051 obtained over 45 days to determine the ultraviolet (UV) light curve variability characteristics and search for correlated UV/X-ray emission. The UV light curve shows variability on all time scales, however with lower fractional rms than the 0.2--10 keV X-rays. On days-weeks timescales the fractional variability of the UV is $F_{\rm var} \sim 8 \%$, and on short ($\sim$~hours) timescales $F_{\rm var} \sim 2 \%$. The \textit{within-observation} excess variance in 4 of the 15 UV observations was found be much higher than the remaining 11. This was caused by large systematic uncertainties in the count rate masking the intrinsic source variance. For the four ``good" observations we fit an unbroken power-law model to the UV power spectra with slope $-2.6 \pm 0.5$. We compute the UV/X-ray Cross-correlation function for the ``good" observations and find a correlation of $\sim 0.5$ at time lag of $\sim 3$ ks, where the UV lags the X-rays. We also compute for the first time the UV/X-ray Cross-spectrum in the range 0--28.5 ks, and find a low coherence and an average time lag of $\sim 3$ ks. Combining the 15 \textit{XMM-Newton} and the \textit{Swift} observations we compute the DCF over $\pm$40 days but are unable to recover a significant correlation. The magnitude and direction of the lag estimate from the 4 ``good" observations indicates a scenario where $\sim 25$ \% of the UV variance is caused by thermal reprocessing of the incident X-ray emission.
\end{abstract}


\section{Introduction}
The broad-band spectral energy distribution (SED) of Active Galactic Nuclei (AGN) requires multiple emission components, including thermal and non-thermal mechanisms \citep{shang11}. The energy output in Seyfert galaxies and quasars is dominated by the ultraviolet (UV) emission, thought to be mostly thermal emission from the inner parts of an accretion disc surrounding the central supermassive black hole (SMBH). The location of the UV emitting region depends on the details of the accretion flow, and these in turn depend on the black hole mass and accretion rate, but is typically $\sim 10-1000 r_g$ (where $r_g = G M_{BH} / c^{2}$ is the gravitational radius). By contrast the X-ray spectrum is usually interpreted as the result of inverse-Compton scattering of soft thermal photons by an optically thin corona of hot electrons in the central few tens of $r_g$ from the black hole (e.g. Haardt \& Maraschi 1991).

The causal connections between these processes are still unclear but can in principle be investigated by studying the time variations in the luminosity across different wavebands. Strong UV and X-ray variability is common to AGN on a wide range of timescales (e.g. \citealt{collin01}), with the most rapid variations seen in X-rays (e.g. \citealt{mushotzky93}). If the emission mechanisms are coupled the UV and X-ray variations should be correlated, in which case the direction and magnitude of time delays should reveal the causal relationship. Two favoured coupling mechanisms are i) Compton up-scattering of UV photons --- produced in the disc --- to X-ray energies in the corona \citep{HaardtMaraschi91}, and ii) thermal reprocessing in the disc of X-ray photons produced in the corona \citep{guilbertrees88}. The interaction timescale for these two processes is approximately the light crossing time between the two emitting regions, and will be in the region of minutes to days for black hole masses $\Mbh \sim10^{6}-10^{8} \Msun$.

Both emission regions are also likely to be correlated at some level if they are both modulated by their local accretion rate, which varies as accretion rate fluctuations propagate through the flow (e.g. \citealt{arevalouttley06}). In a standard accretion disk the timescale for propagation of fluctuations between the two emission regions is governed by the viscous timescale of the disc. For $\Mbh \sim10^{6}-10^{8} \Msun$ this timescale will be in the region of weeks to years \citep{czerny06}. If any of these processes are significant, their effects should be apparent from time series analysis of light curves from both wavebands. A combination of these processes occurring at the same time could make individual reprocessing models harder to detect.
 
Studies of correlations between variations in different wavebands are a potentially powerful tool for investigating the connections between different emission mechanisms (see \citealt{uttley06} for a short review). X-ray/optical correlations on long timescales have been seen in radio-quiet AGN (e.g. \citealt{uttley03}; \citealt{arevalo08}; \citealt{arevalo09}; \citealt{breedt09}). Together with the optical-optical lags (e.g. \citealt{cackett07}) they imply that a combination of accretion fluctuations and reprocessing produces much of the optical variability. X-ray/UV correlations have been seen in e.g. \citet{nandra98}; \citet{cameron12}, however, there are currently fewer examples than the X-ray/optical correlation studies.

The target of the present paper -- the low-mass Narrow line Seyfert 1 (NLS1) galaxy NGC 4051 -- has been the subject of several such studies. \citet{done90} obtained $\sim 3$ days of contemporaneous X-ray, UV, optical and IR data,  but found little variability at longer wavelengths despite strong, rapid X-ray variations. Using light curves spanning $\sim 1000$ days in the optical and X-rays, \citet{peterson2000} revealed significant optical variability on longer timescales that appeared to be correlated with the (longer timescale) X-ray variations. \citet{shemmer03} and \citet{breedt10} also found a significant X-ray/optical correlation, the latter using $\sim 5000$ days of monitoring data. On shorter timescales \citet{mason02} and \citet{smithvaughan07} used $\sim 2$ day {\it XMM-Newton} observations to search for X-ray/UV correlations, with inconclusive results.

The rest of this paper is organised as follows. Section \ref{sect:obsdata} discusses the observations of NGC 4051 and extraction of the X-ray and UV light curves. Section \ref{sect:var} gives an analysis of the variability amplitudes and UV power spectral density, cross-correlations are discussed in section \ref{sect:cor}, and the implications of these results are discussed in section \ref{sect:sum}.

\section{Observations and Data Reduction}
\label{sect:obsdata}
\subsection{Target and observations}
NGC 4051 is a nearby ($z = 0.0023$) NLS1 galaxy with black hole mass, $\Mbh \approx 1.7 \pm 0.5 \times 10^{6} \Msun$ \citep{denney09}, at a Tully-Fisher distance, $D \approx 15.2$ Mpc \citep{russell02}.

Simultaneous, or quasi-simultaneous, observations of the two bands are required in order to search for correlated variability. The \textit{XMM-Newton} and \textit{Swift} observatories are suited for this task. The three EPIC X-ray detectors (pn; \citealt{struder01}, MOS1/2; \citealt{turner01}) and co-aligned Optical Monitor (OM; \citealt{mason01a}) makes \textit{XMM-Newton} an ideal instrument for probing the X-ray to UV correlation on short time scales. The rapid response and flexible scheduling of \textit{Swift} (\citealt{gehrels04}), with the X-ray Telescope (XRT; \citealt{burrows04}) and co-aligned Ultraviolet/Optical Telescope (UVOT; \citealt{roming05}), allow for X-ray and UV monitoring on longer timescales.

NGC 4051 was observed by \textit{XMM-Newton} in 15 separate observations over a period of 45 days during May-June 2009. Each observation lasted $\sim$ 40 ks giving a total of $\sim$ 580 ks of usable data. Individual observation details are listed in table~\ref{obs_tab}. We make use of the EPIC-pn detector in small window mode, with an $\sim 4 \times 4$ arcmin field-of-view. With roughly the same start and end time as the OM, this gives a total of $\sim$ 530 ks of simultaneous UV and X-ray data.

All of the OM observations were taken in \textit{Imaging Mode} in the UVW1 filter (central wavelength 2910 \AA) with $2 \times 2$ pixel binning. This gives a field-of-view of $8 \times 8$ arcmin and a spatial resolution of 0.95 arcsec/pixel. The exposure length varied between 1000--1400 sec from observation to observation, giving $\sim$ 30 exposures per observation and a total of 406 across all 15 observations.

To compliment the \textit{XMM-Newton} data set, 51 \textit{Swift} ToO observations were made covering the same epoch. The observations are separated by $\sim$ 1 day and typically $\sim$ 1.5 ks long, with 1--3 UVOT exposures per observation, giving a total of 71 frames. The UVOT exposures were taken in the \textit{uvw}1 filter, which has approximately the same bandpass as the OM UVW1, with a field-of-view of $17 \times 17$ arcmin and pixel scale of 0.5 arcsec. The XRT has a $23 \times 23$ arcmin field-of-view with a pixel scale of 0.236 arcsec and covers the energy range 0.2--10.0 keV, similar to EPIC-pn.

\begin{table}
\caption{\textit{XMM-Nexton} observation summary. The columns list (1) the spacecraft revolution number, (2) the start date of the onservation, (3) the EPIC-pn observation duration, (4) the OM observation duration, (5) the number of OM images in each observation.}
\begin{tabular}{c c c c c}
\hline
XMM & Observation & EPIC-pn & OM & No. OM\\
rev. no. & Date & On time & On time & images\\
& [Y-M-D] & [s] & [s] & \\
\hline
1721 & 2009-05-03 & 45717 & 45105 & 33\\
1722 & 2009-05-05 & 45645 & 45103 & 33\\
1724 & 2009-05-09 & 45548 & 45003 & 33\\
1725 & 2009-05-11 & 45447 & 44903 & 33\\
1727 & 2009-05-15 & 32644 & 31340 & 21\\
1728 & 2009-05-17 & 42433 & 37367 & 26\\
1729 & 2009-05-19 & 41813 & 41267 & 26\\
1730 & 2009-05-21 & 41936 & 40894 & 26\\
1733 & 2009-05-27 & 44919 & 39168 & 22\\
1734 & 2009-05-29 & 43726 & 43182 & 27\\
1736 & 2009-06-02 & 44946 & 44164 & 22\\
1737 & 2009-06-04 & 39756 & 34574 & 23\\
1739 & 2009-06-08 & 43545 & 43000 & 27\\
1740 & 2009-06-10 & 44453 & 43909 & 28\\
1743 & 2009-06-16 & 42717 & 42116 & 26\\
\hline
\end{tabular}
\label{obs_tab}
\end{table}

\subsection{OM light curves}
\label{sect:om_lightcurves}
\begin{figure}
\includegraphics[width=0.48\textwidth]{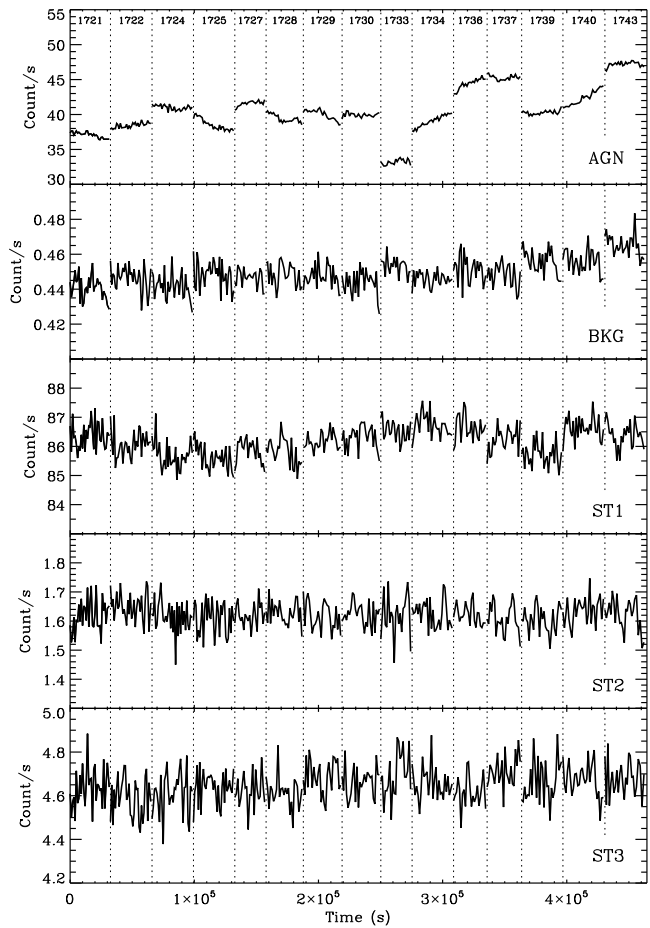}
\caption{Concatenated light curves for the \textit{XMM} OM sources. From top to bottom panel they are central nucleus, background region and field stars 1, 2 and 3 respectively.}
\label{fig:om_ltcrvs}
\end{figure}

The Observation Data Files (ODFs) for our target were extracted from the XMM-Newton Science Archive (XSA) and processed using \textit{XMM-Newton} Science Analysis System (SAS v11.0.0) routine \textsc{omichain}. Custom made \textsc{idl}\footnote{http://www.exelisvis.com} scripts were made to perform source photometry and apply instrumental corrections. Source counts were extracted in a 6 arcsec radius aperture for the galaxy nucleus and 3 field stars present in the images. Background counts were extracted in a 30 arc second radius aperture, placed in a region away from the host galaxy and field stars. Accurate count rates from aperture photometry of the OM images can only be produced once five instrumental corrections have been applied \citep{mason01a}. These are, in order of application to the extracted counts: the point spread function (PSF1), coincidence loss (CL), CCD dead-time (DT), the UV point spread function (PSF2) and time-dependent sensitivity degradation (TDS) corrections. The concatenated background subtracted light curves for the \textit{XMM} OM sources and background region are shown in Fig.~\ref{fig:om_ltcrvs}.

In the UVW1 filter there will be a significant contribution to the observed nuclear light from the host galaxy. This should be constant (to within the random and systematic errors of the aperture photometry) and so we have not tried to remove it, but as such it should not affect the PSD or correlation analysis in any important way.

As a test of the background subtraction and photometry procedure we tested for (zero lag) correlations between the background subtracted light curves of the sources and a second background region for all 15 observations. For $\sim$~400 data points a Pearson linear correlation coefficient $r \geq 0.13$ indicates a weak but statistically significant correlation ($p < 0.01$). We find no significant correlation between each of the background subtracted sources, but in the source vs background tests, values of \textit{r} up to 0.3 are observed. The strength of this correlation is also observed to change between 0.0 and 0.3 when using a different background region. The mean correlation coefficient between source vs source and source vs background for individual observations is very low ($-0.1 \la r \la 0.1$). This indicates that the correlation is caused by changes in the background over the course of the 15 observations. We are cautious of this fact during the rest of the analysis and use source light curves subtracted using various background regions. We find that the choice of background region has no effect on any subsequent analysis.

\subsection{EPIC-pn light curves and spectra}
The EPIC-pn light curves used in this analysis are the same as those used in \citealt{vaughan11a}. The raw EPIC-pn data were processed from the ODFs using the SAS (v11.0.0). Events lists were filtered using \textsc{pattern} 0--4, \textsc{flag} $=0$ and were visually inspected for background flaring. Light curves were extracted from the filtered events files using a source aperture of radius 35 arcsec and a non-overlapping larger background region on the same chip. Light curves were extracted with bin size of $\Delta t = 5$ s and an energy range of 0.2--10.0 keV. The background subtracted EPIC-pn light curves are shown in the bottom panel of Fig.~\ref{fig:uv_x_ltcrvs}. Spectra were extracted and binned to a minimum of 25 counts per bin. Response files were created using \textsc{rmfgen} v1.55.2 and \textsc{afrgen} v.1.77.4.

\subsection{UVOT and XRT light curves}
Visual inspection of the 71 UVOT exposures revealed considerable target movement in 40, leaving 31 usable frames. The exposure time varied slightly across the observations but is typically $\sim$ 500 s. The HEAsoft (v6.10) tasks \textsc{uvotsource} was used to extract source counts from the galaxy nucleus and the same field stars used in the OM photometry. The optimum extraction radius is 12.5 pixels ($\sim$ 6 arc seconds) for the \textit{uvw}1 filter \citep{poole08}. A 60 pixel radius background aperture fixed in sky co-ordinates in a blank region of the sky was also extracted. Similar to the OM, instrumental corrections --- PSF1, CL, DT, PSF2 and TDS --- must be applied to the counts extracted in each aperture on the UVOT images. \textsc{uvotsource} automatically performs background subtraction and applies these corrections based on the most up-to-date instrument calibration database (CALDB 4.1.2). The background subtracted UVOT light curves are shown in the lower panel of Fig.~\ref{fig:uv_x_ltcrvs}.
We extract XRT counts in the 0.2--10 keV range using the online XRT Products Builder \citep{evans09}. This performs all the necessary processing and provides background subtracted light curves.

\begin{figure*}
\includegraphics[width=0.6\textwidth,angle=90]{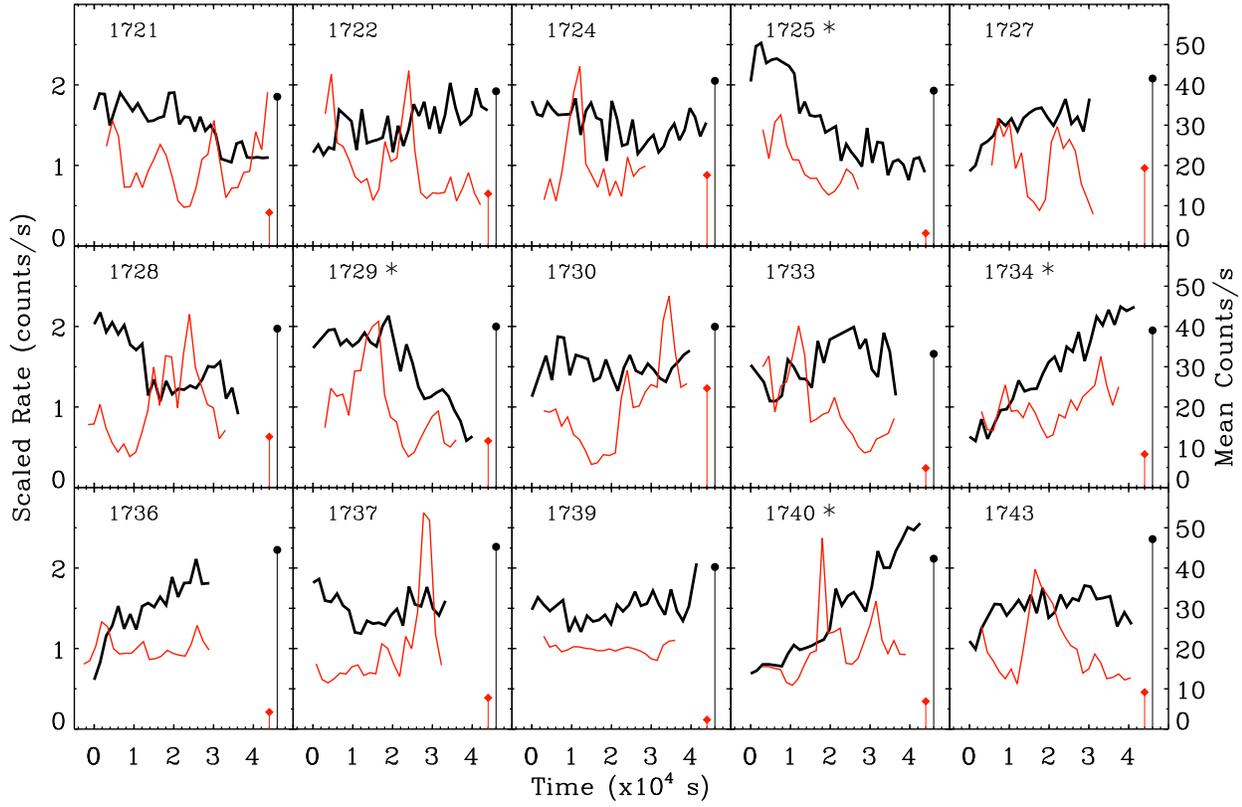}
\caption{Within-observation X-ray and UV variability from the 15 \textit{XMM-Newton} observations. To better represent the light curves the OM (thick black) has been scaled to some arbitrary value and shifted to a mean of 1.5 ct/s. The EPIC-pn (thin red) has been normalised to a mean rate of 1 ct/s. The circle and diamond markers represent the mean observation count rate, given by the right ordinate, for the OM and EPIC-pn respectively. The four ``good" OM observations are indicated by the asteriks next to the revolution number.}
\label{fig:ltcrv_xmm_sec}
\end{figure*}

\begin{figure*}
\includegraphics[width=0.44\textwidth,angle=90]{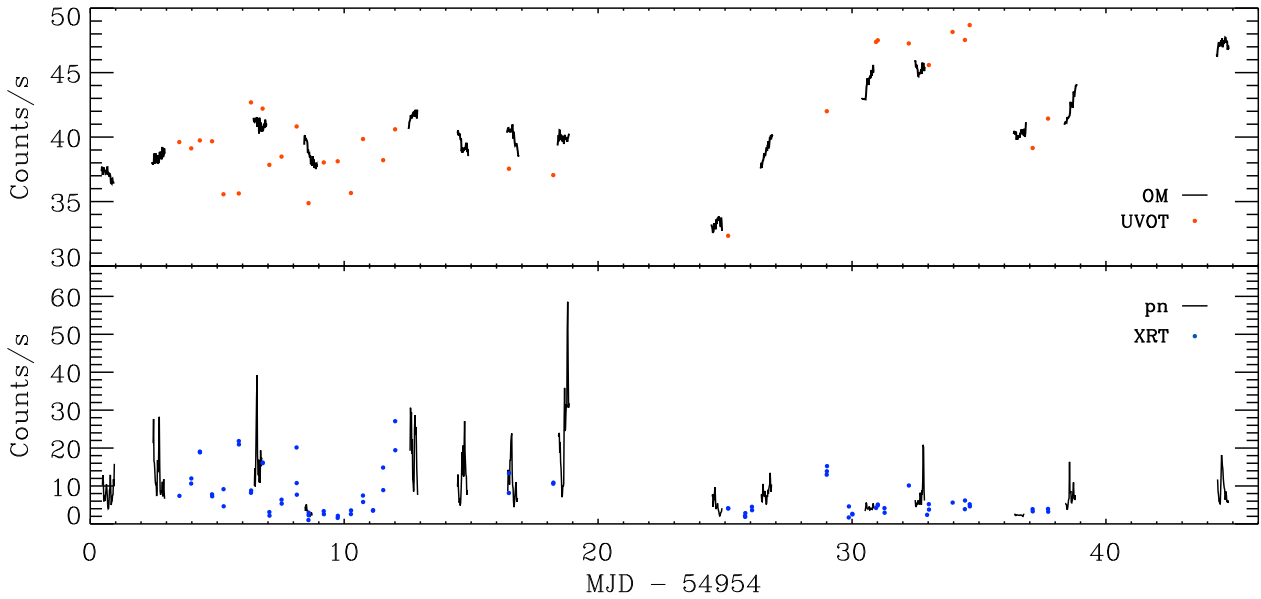}
\caption{Between-observation variability for UV (top) and X-ray (bottom) from \textit{XMM} and \textit{Swift}. In the  top panel the black lines are the OM data and the red circles are the UVOT. In the bottom panel the black lines are the EPIC-pn data and the blue dots the XRT. In both panels the \textit{Swift} count rates have been scaled up to account for different effective areas (see section~\ref{sec:between} for details).}
\label{fig:uv_x_ltcrvs}
\end{figure*}

\section{Data Analysis}
\label{sect:var}
\subsection{Quantifying variability}

In an observed light curve some of the total variance will be intrinsic to the source and some will come from variations in the measurement uncertainties \citep{vaughan03a}. The difference between the two is the `excess variance', and can be used to estimate the intrinsic source variance. The values of the variability estimators for the OM and UVOT sources are listed in table~\ref{tab:quant_var}. The variability statistics calculated over the whole $\sim45$ day {\it XMM-Newton} observation show there is significant variability in NGC 4051, with $F_{\rm var} \approx 8\%$. On timescales within each observation, the  variability is weaker than on long timescales. 

In 11 of the {\it XMM-Newton} observations the excess variance in the nucleus and star 1 values are almost identical. This indicates that there is a ``floor" in the excess variance, that isn't accounted for by the errors. We refer to this as a sytematic error but are unable to account for it. In the 11 observations this systematic error is too large to detect any intrinsic variability from the nucleus. The variability estimators in table~\ref{tab:quant_var} show the nucleus in the remaining 4 ``good" observations clearly posses significant variability compared to the field stars, and as such will be treated separately form  the the ``poor" 11 observations in the correlation analysis in section \ref{sect:cor}. The ``good" UV observations are revolutions 1725; 1729; 1734; and 1740, and are indicated by an asteriks in Fig.~\ref{fig:uv_x_ltcrvs}.

\begin{table}
 \caption{Quantifying source UV variability. The columns give the following information: (1) source name, (2) mean count rate, (3) sample standard deviation, (4) excess variance, (5) fractional, excess rms amplitude ($F_{\rm var}$). The upper table shows the values obtained by combining all $15$ {\it XMM-Newton} OM observations. The second and third tables show the values obtained by averaging the results from each of the $4$ ``good" and $11$ ``poor" {\it XMM-Newton} OM observations respectively. The lower table shows the results from the {\it Swift} UVOT observations.}
 \label{tab:quant_var}
 \begin{tabular}{@{}lccccc}
 \hline
Object & Mean rate & $\sigma$ & $\sigma^{2}_{\rmn{XS}}$ & $\rmn{F}_{\rmn{var}}$ \\
 & ct/s & ct/s & $[\rmn{ct/s}]^{2}$ & percent \\
 \hline
 \multicolumn{5}{|c|}{{\it XMM-Newton} total} \\
Nucleus & 40.5 & 3.19 & 10.1 & 7.9 \\
Star 1 & 86.2 & 0.55 & 0.24 & 0.6 \\
Star 2 & 1.6 & 0.05 & 0.0005 & 1.5 \\
Star 3 & 4.6 & 0.09 & 0.003 & 1.2 \\
Background & 0.4 & 0.14 & 0.015 &  2.6 \\
 \hline
 \multicolumn{5}{|c|}{{\it XMM-Newton} ``good" 4 observation averages} \\
Nucleus & 40.0 & 0.84 & 0.69 & 2.1 \\
Star 1 & 86.3 & 0.37 & 0.09 & 0.3 \\
Star 2 & 1.6 & 0.05 & 0.0005 & 0.2 \\
Star 3 & 4.7 & 0.08 & 0.001 & 0.6 \\
Background & 0.4 & 0.006 & 0.005 & 1.0 \\
 \hline
 \multicolumn{5}{|c|}{{\it XMM-Newton} ``poor" 11 observation averages} \\
Nucleus & 40.7 & 0.40 & 0.14 & 0.9 \\
Star 1 & 86.2 & 0.42 & 0.13 & 0.4 \\
Star 2 & 1.6 & 0.05 & 0.0004 & 0.1 \\
Star 3 & 4.6 & 0.08 & 0.003 & 0.9 \\
Background & 0.4 & 0.007 & 0.007 & 1.8 \\
 \hline
 \multicolumn{5}{|c|}{{\it Swift} total} \\
Nucleus & 40.1 & 4.39 & 19.2 & 10.8 \\
Star 1 & 48.5 & 1.06 & 1.01 & 0.02 \\
Star 2 & 1.1 & 0.04 & -0.001 & 0.02 \\
Star 3 & 3.1 & 0.09 & 0.002 & 0.01 \\
\hline
\end{tabular}
\end{table}

\subsection{The UV Power Spectrum}
The power spectral density (PSD) describes the amount of variability power present in the light curve (mean squared amplitude) as a function of temporal frequency. The PSD of the X-ray light curves is discussed by \citet{vaughan11a}. The UV power spectrum was estimated from the 15 individual OM observations using standard methods (e.g \citealt{vanderklis89}). A 30 ks segment (equal to the shortest observation length) was taken from each observation. Within each {\it XMM-Newton} observation the individual OM exposures are approximately evenly sampled in time, although the exposure times do differ between observations (from $1200$ to $1500$ s). The basic periodogram requires evenly sampled data, and so we interpolated all OM data onto a grid evenly sampled at $\Delta t = 1500$ s -- the smoothness of the OM light curves means that linear interpolation should not affect the shape of the time series in any significant manner. The observed power spectrum may be distorted by leakage of power from low frequencies to higher frequencies (van der Klis 1989; \citealt{uttley02a}; Vaughan et al. 2003). This can bias the data such that the observed spectrum resembles an $\alpha = 2$ power law even if the true power spectrum is somewhat steeper.

\begin{figure}
\includegraphics[width=0.32\textwidth,angle=90]{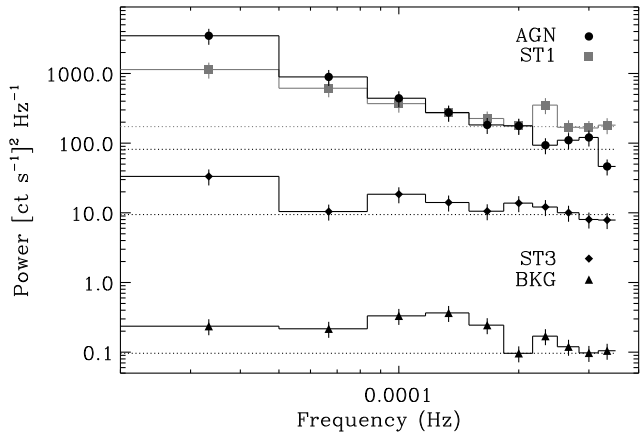}
\caption{UV power spectrum in absolute units for the background subtracted OM sources, produced from the 15 \textit{XMM} observations. The circles, squares, diamonds represent NGC 4051, star 1 and star 3 respectively. The triangles are the power spectrum estimate for the background light curves subtracted by a background region on the opposite side of the CCD, see text for details. The dotted lines are the Poisson noise estimates for each source.}
\label{fig:uv_psd_src}
\end{figure}

Figure~\ref{fig:uv_psd_src} shows the power spectrum for the NGC 4051 nucleus, star 1, star 3 and the background for the 15 {\it XMM-Newton} OM observations. Periodograms were computed with absolute normalisation and the Poisson noise level is estimated using the formula in Vaughan et al. (2003); Appendix A. The background power spectrum is computed for the light curve from the background region (the same background region used in the source background subtraction) subtracted by a second background region on the opposite side of the CCD. In all sources, some power above the noise level is present. This is most likely the result of the background subtraction issues described in section~\ref{sect:om_lightcurves}. Star 1 shows a similar red-noise slope, albeit with less power, to the nucleus. A cross-correlation test (see section~\ref{sect:cor}) between the nucleus and star 1 revealed no significant correlation between the two sources. This indicates that the variations in star 1 are either intrinsic to the star or caused by the problems in the background subtraction.  

\begin{figure}
\includegraphics[width=0.34\textwidth,angle=90]{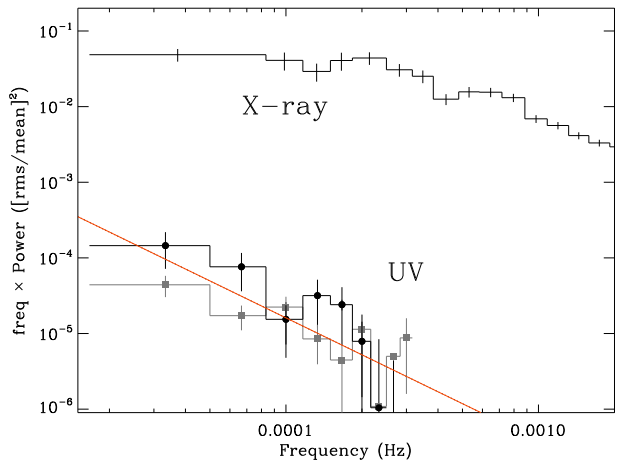}
\caption{UV power spectrum produced from the \textit{XMM} OM observations. The black circles are the average of the 4 ``good" periodograms and the grey squares are the average of the 11 ``poor" periodograms, prior to subtraction of Poisson noise. The red solid line is the power law fit to the ``good" data. The upper solid black line is the Poisson noise subtracted X-ray ($0.2-10$\,keV) power spectrum from \citealt{vaughan11a}.}
\label{fig:uv_powspec}
\end{figure}

Figure~\ref{fig:uv_powspec} shows the resulting NGC 4051 power spectrum from the ``good" and ``poor" OM data, and a two component model is fit to the ``good" data. Periodograms were computed with fractional rms normalisation (Vaughan et al. 2003; Appendix A). We did not subtract the expected contribution from Poisson noise but instead included this in the model fitting. The simple model comprises a power law plus constant to account for the Poisson fluctuations in the count rate: $P(\nu) = A\nu^{-\alpha} + P_N$ (where $\nu$ is the temporal frequency, $A$ is a normalisation term, $\alpha$ is the power law index and $P_N$ is the power density due to Poisson noise). This was fitted to the data using \textsc{xspec} v12.6.0 \citep{arnaud96}. The $P_N$ level was allowed to vary freely. Using a ${\Large \chi}^{2}$ statistic the best fit to the data is found to have $\alpha = 2.62 \pm 0.48$ and $P_N = 0.059 \pm 0.023$, with ${\Large \chi}^{2}$ = 3.7 for 7 degrees of freedom (\textit{dof}). The Poisson noise level can also be estimated from the formula given in \citet{vaughan03a} which we compute to be $P_N = 0.05$, in agreement with the value derived from the PSD. As expected, fitting the model assuming a fixed $P_N=0.05$ value gave consistent results for the index parameter ($\alpha = 2.50 \pm 0.35$), with ${\Large \chi}^{2}$ = 3.9 for 8 \textit{dof}. Errors on the model parameters correspond to a 90 per cent confidence level for each interesting parameter (i.e. a ${\Delta {\Large \chi}^{2} = 2.7}$ criterion). For comparison, the X-ray PSD from \citet{vaughan11a} is plotted in Fig.~\ref{fig:uv_powspec}.

\section{Correlation Analysis}
\label{sect:cor}
In this section we discuss tests for X-ray/UV correlations on short timescales within each observation (\textit{within -observations}) and longer timescale (\textit{between-observations}). Treated individually, each of the 15 \textit{XMM} observations allowed us to probe time scales of $\sim2-40$ ks. Combining the \textit{XMM} and \textit{Swift} light curves allowed us to look for any long term trends in correlation between the two bands over the $\sim45$ days.

\subsection{Within-observation correlations}
Standard time series analysis methods (c.f \citealt{boxjen76}; \citealt{priestley81}) require the two light curves to be simultaneous and evenly sampled. This requirement is complicated by the the OM and EPIC-pn not always starting and ending at the same time, the irregular sampling of the OM, the read out time of the OM CCD, and any bad OM exposures. Where the two light curves are simultaneous we linearly interpolate the OM onto an uniformly-sampled regular grid. Given that the OM light curves vary smoothly within each observation, linear interpolation should not have a significant effect on the intrinsic variability. The EPIC-pn X-ray counts are re-binned to be contiguous and simultaneous with the OM bins by taking the average count rate within the new bin width. We chose a bin width of 1500 s to be consistent with the mean sampling rate of the OM (1502 s) across the 15 \textit{XMM} observations. The simultaneous light curve lengths range from 28.5-43.5 ks.

\subsubsection{The Correlation Function}
\begin{figure}
\includegraphics[width=0.46\textwidth]{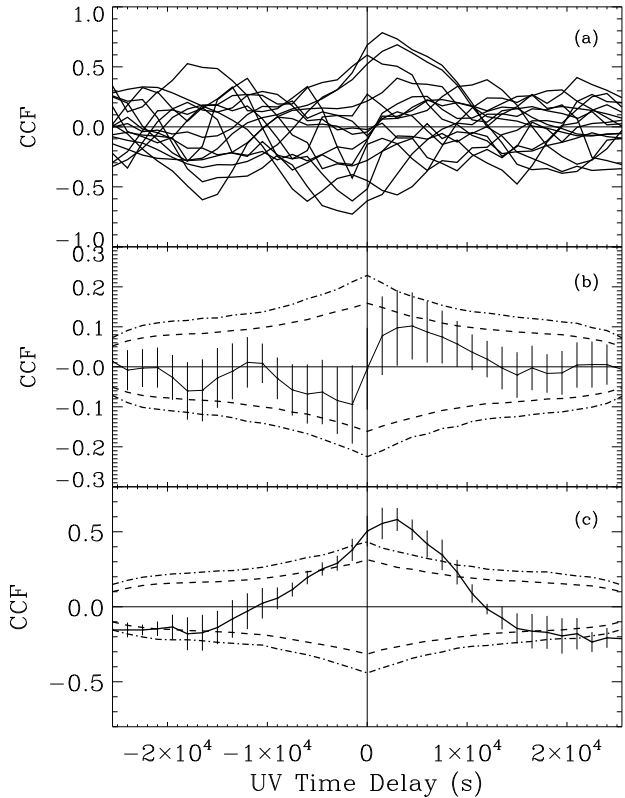}
\caption{The Cross-correlation function (CCF) for the 15 \textit{XMM} observations. The top panel is the individual CCF for each observation, the middle panel is the average CCF of all 15 observations, whilst the bottom panel is the average CCF for the 4 ``good" observations. The dashed and dot-dashed lines are the $95\%$ and $99\%$ confidence intervals respectively, calculated individually for each data subset. Conficence intervals were estimated using Monte Carlo simulations of light curves assuming no correlation (see Appendix~\ref{ap:sim})}
\label{fig:ccf}
\end{figure}

The cross-correlation function (CCF) is a standard tool for measuring the degree of correlation between two evenly sampled time series ($x_{\rmn{t}}, y_{\rmn{t}}$) as a function of time-lag (c.f \citealt{boxjen76}; \citealt{priestley81}). We estimated the CCF for each \textit{XMM-Newton} observation individually using the \textsc{idl} function \textsc{c\_correlate}, shown in panel \textit{a} of Fig.~\ref{fig:ccf}, where a positive lag in the plot indicates the UV variations are lagging those of the X-rays. A large spread in the CCF value for any computed time lag is seen. Panel \textit{b} of Fig.~\ref{fig:ccf} shows the average CCF for all 15 \textit{XMM} observations. The strongest feature is the peak in the CCF around $\sim$ 4.5 ks, although with a correlation of $\sim$ 0.1, which falls within the confidence intervals. Confidence intervals on the average CCF are estimated using Monte Carlo simulations, where the $95\%$ and $99\%$ confidence intervals are shown in Fig.~\ref{fig:ccf}. These show the expected range of CCF values under the assumption that the X-ray and UV processes are independent, i.e. in the absence of a real correlation. The full details of these simulations is given in Appendix~\ref{ap:sim}. The error on the average CCF is given by the standard error for $N$ observations at each time-lag \textit{t}.

Panel \textit{c} in Fig.~\ref{fig:ccf} shows the average CCF plot for the 4 ``good" observations (see section~\ref{sect:var}). A distinct broad peak can be seen around $\sim$ 3 ks with a correlation of $\sim$ 0.5, which lies outside the $99\%$ confidence interval. Confidence intervals are calculated the same as above except 4 simulated light curves are averaged over in each CCF estimate. The small error bars on the averaged ``good" CCF shows there is little scatter in the individual CCF measurements.

A correlation between optical light curves and X-ray photon index has been detected in some sources (e.g. \citealt{nandra2000}), despite there being a weak correlation between the optical and X-ray light curves. We therefore cross-correlated the UV light curves with the 0.7--2/2--10 keV hardness ratio (a proxy for photon index) but find a CCF shape similar to that between the UV and X-ray light curves. This is most likely due to the X-ray spectral shape changes being strongly correlated with the overall X-ray flux.

\subsubsection{The Cross-Spectrum}
\begin{figure}
\includegraphics[width=0.44\textwidth]{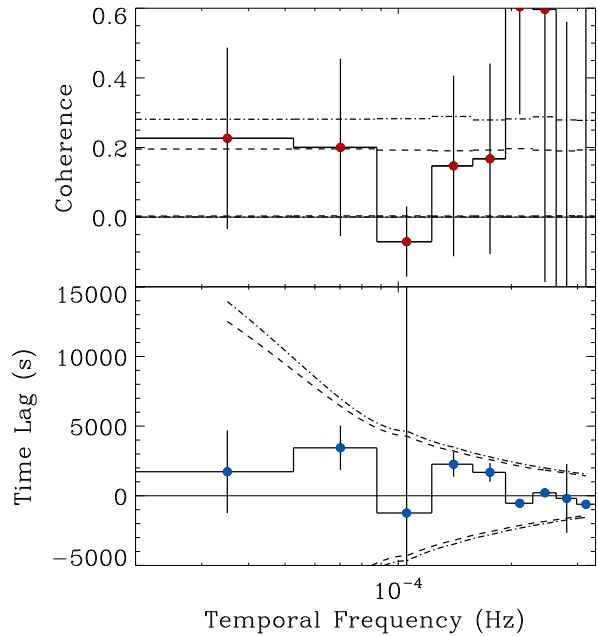}
\caption{The Coherence and Phase-lag parts of the Cross-spectrum for the combined 4 ``good" \textit{XMM} observations. The dashed and dot-dashed lines are the $95\%$ and $99\%$ confidence intervals respectively. Conficence intervals were estimated using Monte Carlo simulations of light curves assuming no correlation (see Appendix~\ref{ap:sim})}
\label{fig:xs}
\end{figure}

The cross-spectrum is the Fourier transform of the CCF (\citealt{boxjen76}, \citealt{priestley81}), and has been widely used in analysing X-ray light curves from X-ray binaries (e.g \citealt{vaughannowak97}; \citealt{nowak99}; \citealt{miyamoto89}) and more recently from AGN (e.g. \citealt{fabian09}). It contains the same information as the CCF but represents the time-lags and strength of correlation in terms of phase difference and coherence as a function of temporal (Fourier) frequency. The phase lag $\Delta \phi$ can be expressed as a time-lag at a given frequency $\nu$: $\tau =  \Delta \phi / 2 \pi \nu$. Under quite general conditions the phase delay estimates are approximately independent at each frequency; by contrast, adjacent values of the CCF tend to be correlated due to the autocorrelation of the individual time series. Here we have estimated the cross spectrum using the ``good" OM data, except that the segments have been trimmed to equal length (28.5 ks), corresponding to the shortest simultaneous light curve. The resulting coherence and phase parts of the cross-spectrum are shown in Fig.~\ref{fig:xs}. Errors were estimated using standard formulae \citep{vaughannowak97}, and confidence intervals were estimated using simulated light curves (see appendix~\ref{ap:sim} for details).

The coherence between the two bands is found to be low ($\leq 0.2$) at all frequencies. The average time-lag of the lowest 5 frequency bins is $\sim$ 3 ks, consistent with what is seen in the CCF. A low coherence means that the errors on the time delay estimates are most likely underestimated using standard formulae, which can increase the apparent significance of lags when the errors are estimated using the standard formula and the intrinsic coherence is very low (e.g. Bendat \& Piersol 1986). The cross spectrum is also computed for the combined 15 \textit{XMM} observations which gives a coherence consistent with zero for each frequency bin, and the average time-lag in the lowest 5 bins is consistent with that found with the 4 ``good" observations.


\subsubsection{Pre-processing the light curves}
The OM light curves tend to be dominated by slow, quasi-linear trends, and these can affect the CCF estimation (Welsh 1999). We have repeated the CCF and cross-spectrum analysis after `end-matching' the OM light curves (i.e. removing a linear trend such that the first and last points are level - see \citealt{fougere85}). This `end-matching' removes, to a large extent, linear trends from the data, and alleviates the problem caused by circularity of the Fourier transform when estimating the cross spectrum.

We end-matched the \textit{XMM} OM light curves individually and computed the CCF and cross-spectrum using the same X-ray light curves as before. The CCF for both the ``good" and \textit{all} the data remains mostly unchanged. In the cross-spectrum the phase-lag follows the same distribution and the coherence remains low in both cases. As this reanalysis did not substantially alter the results we do not show the CCF and cross-spectrum plots here.


\subsection{Between-observation correlations}
\label{sec:between}
Given the extended period and sampling of the \textit{XMM} and \textit{Swift} observations, we are able to search for possible correlations and time lags on longer time scales. The {\it XMM-Newton} data (EPIC-pn and OM) and {\it Swift} data (XRT and UVOT) were first treated separately, then combined to produce one X-ray and one UV light curve. In either case the time sampling between observations is highly uneven, and so the Discrete Correlation Function (DCF; \citet{edelson_88_dcf}) was used to estimate the CCF. We use all the {\it XMM-Newton} OM data in this part of the analysis, as the variations within each OM observation will have little affect on the DCF.

\begin{figure}
\includegraphics[width=0.46\textwidth]{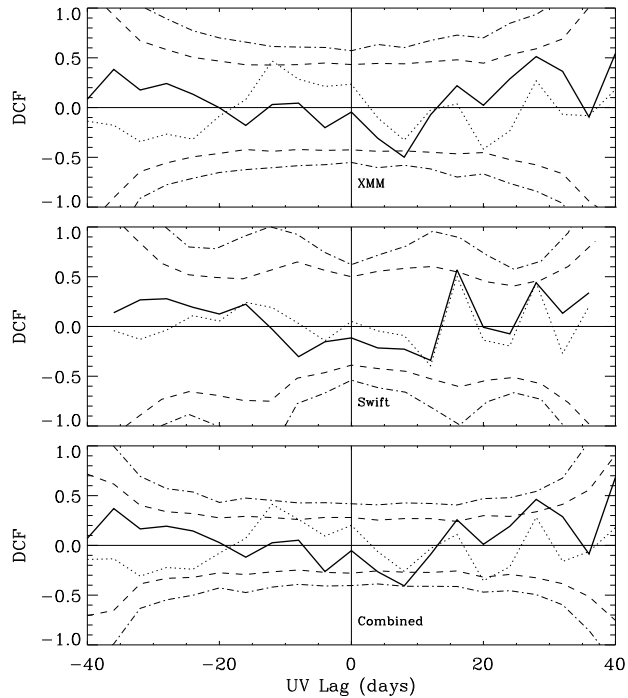}
\caption{The discrete correlation function (DCF) for the between-observations (solid black line). The dotted black line is the DCF for the end-matched data. The dashed and dot-dashed lines represent the 90$\%$ and 95$\%$ confidence intervals respectively.}
\label{fig:dcf_plot}
\end{figure}

For the \textit{XMM} dataset, the midpoint of each original OM exposure was used. The EPIC-pn data were binned to be contiguous from the start of each revolution, with a bin size of 1500 s to be consistent with the mean OM sampling rate.  A 10 ks DCF bin width is adopted to be consistent with the mean OM sampling rate over the extended observation. Although the OM exposure length varied between 1200--1500 s from revolution to revolution, we treat the source count rate in each exposure as a representative of the average count rate. As the source varies smoothly in the UVW1 filter we do not expect this to have any effect on the shape of the DCF. We test this by computing the DCF using the OM data that was binned onto a 1500 s even grid, and find no change in the shape of the DCF. The DCF for the \textit{XMM} dataset for the range $-40<\rmn{lag}<+40$ days is shown in Fig.~\ref{fig:dcf_plot}, where a positive lag means the UV are lagging the X-rays. Some peaks can be seen in the DCF but all lie within the confidence intervals. The peaks are most likely the result of the DCF binning used, combined with the underlying shape of the uncorrelated red-noise light curves.

For the \textit{Swift} dataset, the UVOT exposures from each snapshot were used to represent the mean source count rate in the middle of each exposure bin. Again the exposure lengths varied from $\sim$ 300--800 s with a mean of $\sim$ 500 s, but due to the steepness of the red-noise power spectrum this will have no effect on the shape of the DCF as long as the DCF bin size is much greater than the mean UVOT exposure length. The XRT counts are taken from each snapshot and have a mean exposure length of $\sim$ 500 s. The DCF for the \textit{Swift} data is plotted in Fig.~\ref{fig:dcf_plot}. The plotted $95\%$ and $99\%$ confidence intervals are calculated using simulated light curves following the method outlined in appendix~\ref{ap:sim}. 

To make the most of the observational coverage we combine the \textit{XMM} and \textit{Swift} datasets and recompute the DCF. As the effective areas of the UV and X-ray instruments on either telescope are not identical the count rates from one telescope need to be scaled before the DCF can be computed. We estimate this scale factor using the 3 occasions the observations overlap and find OM $\approx 1.1 \times$UVOT, and EPIC-pn $\approx 15 \times$XRT. The scaling factor for the X-ray cameras is consistent with that calculated by WebPIMMS\footnote{http://ledas-www.star.le.ac.uk/pimms/w3p/w3pimms.html}. No reference for a scaling factor between the UV cameras could be found, but we find the choice of scaling factor within the range $\sim0.5-1.5$ has no effect on the shape of the DCF.

In Fig.~\ref{fig:uv_x_ltcrvs} it can be seen that the UV light curves show a gradual increase in counts over the extended observation. To account for any underlying long-term trends in the UV variability we `end-match' the overall UV light curve and recompute the DCF for the individual and combined datasets. These are shown as the dotted black lines in Fig.~\ref{fig:dcf_plot}.

\section{Discussion}
\label{sect:sum}
Using UV and X-ray data from \textit{XMM-Newton} and \textit{Swift} we have analysed the light curves of NLS1 galaxy NGC 4051 to search for correlations in the variability between the two bands. UV variability is detected on short and long time scales, however the fractional rms amplitude is smaller than that in the X-rays. On days-weeks timescales the fractional variability of the UV is $F_{\rm var} \sim 8 \%$, and on short ($\sim$~hours) timescales $F_{\rm var} \sim 2 \%$ (from the ``good" OM observations).

The excess variance in 4 of the 15 \textit{XMM-Newton} OM observations is found to be considerably greater than the remaining 11. The ``poor" 11 observations show there is a ``floor" to the excess variance that isn't accounted for in the errors. In the ``poor" observations any intrinsic source variations are masked by the errors, and inclusion of these observations will weaken the detection of any correlated emission. Although 4 out of the 15 observations is a relatively small subset, we find the variability statistics of the 4 ``good" observations clearly very different from the remaining 11 ``poor". The similarity in the overall shape of the CCF for the 4 ``good" observations is hard to explain as arising by chance if the they were all representative of uncorrelated emission. Nevertheless, the interpretation of the time lag is still treated with some caution.

Analysis of the UV power spectral density reveals a red-noise light curve with a power-law slope of index $\alpha = 2.62 \pm 0.48$ for the 4 ``good" OM observations. We searched for correlations between the two bands on time scales up to $\sim$ 40 ks, treating all the \textit{XMM-Newton} and the 4 ``good" observations separately. The CCF for the 4 good observations revealed a significant peak of $\sim$ 0.5 at a lag of $\sim$ 3 ks. Using all 15 \textit{XMM-Newton} observations the CCF revealed a weak correlation ($\sim$ 0.1) with a peak at $\sim$ 3 ks. The cross-spectrum showed the lowest 5 frequency bins to have a low mean coherence of $\sim$ 0.2 and a mean phase-lag of $\sim$ 3 ks in both cases. Combining the \textit{XMM-Newton} and \textit{Swift} datasets we searched for correlated emission on timescales up to 40 days and find no significant correlations. The lag significance in the above results was estimated using simulated light curves. A correlation coefficient of $r \sim 0.5$ means the amount of UV variance prdictable from the X-ray varince is $r^{2} \sim 0.25$. As the coherence is a ``square" quantity, this value is consistent with the $\sim$ 0.2 form the coherence.

From a $\sim$ 12 yr monitoring campaign on NGC 4051 using ground based optical photometry \citet{breedt10} estimated the PSD in the frequency range $\sim10^{-8}-10^{-3} \rmn{Hz}$. In their Fig. 4 they fit an unbroken power law to the PSD with $\alpha = 1.4_{-0.2}^{+0.6}$. In the paper \citet{breedt10} fitted a single-bend power-law model to the PSD, as is observed for X-rays (e.g. \citealt{mchardy04}, \citealt{vaughan11a}). Whilst they do not rule out the single-bend model, they find their data is more consistent with an unbroken power law. Their Fig 5 shows the acceptance probabilities for the single-bend model as a function of high frequency slope $\alpha_{\rmn{H}}$ and bend frequency $\nu_{\rmn{B}}$. Taking our value of $\alpha \approx 2.5$ as the high frequency slope this would give a break frequency $\nu_{\rmn{B}} \approx 10^{-6.5}$ Hz. Our PSD is better constrained in the high-frequency range ($\sim10^{-5}-10^{-3} \rmn{Hz}$) and so the slope value $\alpha = 2.62 \pm 0.48$ is consistent with their single-bend model.

In a sample of 4 AGN using \textit{Kepler data}, \citealt{mushotzky11} estimated power spectral slopes (assuming a single power law model) of $\sim$ 2.6--3.3 to the optical PSD in the $\sim10^{-6.5}-10^{-3.5} \rmn{Hz}$ range. They do not attempt to fit a single-bend power-law model to their PSDs, but the break frequency for the larger black hole masses ($\sim10^{7} \Msun$) in their sample would likely occur at lower frequencies than they estimate in the PSD.

The X-ray PSD in Fig.~\ref{fig:uv_powspec} shows orders of magnitude more variability power than the UV. This is consistent with what is seen in optical PSDs, where the high-frequency power is much less than in the X-rays, although the amplitudes can be similar (or even greater) at low frequencies (e.g, NGC 3783, \citealt{arevalo09}). A study with simultaneous UV and X-ray coverage on longer timescales is still lacking. A bend can be seen in the X-ray PSD at $\sim~2 \times 10^{-4}$ Hz (\citealt{vaughan11a}). If a break was present in the UV PSD, it would be seen to occur at much lower frequencies that of the X-ray due to the radius of UV emission being much greater than that of the X-rays.

To assess whether the observed X-ray variations are significant enough to produce the variations seen in the UV band, we compare the root-mean-square luminosity variations in both bands. If the luminosity variations in the UV band are greater than the luminosity variations in the 0.2--10.0 keV band, then this would in effect rule out the 0.2--10.0 keV X-ray variations being the dominant cause of variations in the UV band. The values in table~\ref{tab:lum} show the integrated X-ray luminosity is greater than in the UVW1 band, and the X-ray luminosity variations  are a factor $\sim$~10 greater. It is worth noting here that the UVW1 filter is very narrow compared to the X-rays. The X-ray band covers a factor of $\sim$~50 in wavelength, the UVW1 band covers only a factor 1.3. This largely explains the apparently low luminosity in the UWV1 compared to X-rays. The ratio of the FWHM to the central wavelength of the UVW1 filter is  620\AA/2910\AA $\approx 0.2$. Table~\ref{tab:lum} gives the rms luminosity for X-ray bands of comparable fractional energy range to the UVW1 filter. The rms luminosity in the narrower X-ray bands is now comparable to that of the UV band, albeit with lower mean luminosity. This shows that, in principle, the X-ray variations could drive variations in the UV band.

\begin{table}
 \caption{UV and X-ray rms luminosity for the 15 \textit{XMM-Newton} observations.}
 \label{tab:lum}
 \begin{tabular}{@{}lcc}
 \hline
Band & $\bar{L}$ & $L_{\rmn{rms}}$ \\
 & $10^{41} \rmn{erg/s}$ & $10^{41} \rmn{erg/s}$\\
 \hline
UVW1 & $3.4$ & 0.3 \\
0.2--10 keV X-ray & $7.3$ & 3.9 \\
1--1.2  keV X-ray & 0.3 & 0.2 \\
5--6    keV X-ray & 0.5 & 0.2 \\
\hline
\end{tabular}
\end{table}


Given the published black hole mass ($\Mbh \approx 1.7 \pm 0.5 \times 10^{6} \Msun$ \citealt{denney09}) it is possible to make predicted lag estimates for each reprocessing scenarios based on standard disc equations to find the distance of the UV emitting. In the Compton up-scattering scenario the lags can be expected to be seen in the $\sim$~1.5--7 ks range for assumed accretion rate as a fraction of Eddington of 0.01--0.1. In the thermal reprocessing scenario the time-lags depend on the luminosity of the X-ray band and are expected to be $\sim$~7 ks. The direction and magnitude of our lag from the ``good" data is consistent with the thermal reprocessing scenario. Although the expected time delay of $\sim$~7 ks is predicted from the toy model, the model assumes that the disc is heated solely from the incident X-rays, which are themselves coming from a radius $r = 0$. Both these assumptions are not likely to be true for a real AGN. An extended corona will increase $R_{\rmn{X}}$ and a viscously heated disc will decrease $R_{\rmn{UV}}$ and hence the light travel time between the two emitting regions. In the propagating accretion rate fluctuation model (\citealt{arevalouttley06}) the timescale of mass flow is dictated by the viscous timescale. This is dependent on the assumed viscosity parameter and scale height of the geometrically thin, optically thick accretion disc (\citealt{czerny06}). We estimate this to be in the region of $\sim$ weeks---years.

Given the quality of the UV data in the 4 ``good" observations, a lag in the region of $\sim$~1.5--7 ks would have manifested itself in the cross-correlation analysis. If the lag estimate from the 4 ``good" observations is to be believed, then crudely speaking $\sim25\%$ of the UV and X-ray variance are correlated on timescales of days. This is consistent with the \citet{breedt10} result, where an optical---X-ray correlation of $\sim30\%$ is reported on timescales of $\sim$ weeks.

\section{Acknowledgements}

WNA acknowledges support from an STFC studentship. This research has made use of NASA’s Astrophysics Data System Bibliographic Services, and the NASA/IPAC Extragalactic Data base (NED) which is operated by the Jet Propulsion Laboratory, California Institute of Technology, under contract with the National Aeronautics and Space Administration. This paper is based on observations obtained with XMM-Newton, an ESA science mission with instruments and contributions directly funded by ESA Member States and the USA (NASA).

\label{lastpage}

\bibliography{bib_instr,bib_var,bib_agn}
\bibliographystyle{mn2e}

\appendix

\section{Simulating light curves}
\label{ap:sim}
The red-noise nature of UV light curves means that individual data points in the light curve are correlated with adjacent points. Confidence intervals were placed on the CCF and cross-spectrum measurements from the 15 \textit{XMM} observations using Monte Carlo simulations of uncorrelated light curves. We used the method of \citet{timmerkonig95} to simulate $10^{4}$ light curves in each band with length 50 days, the same time resolution as the re-binned data ($\Delta t = 1500 s$) and appropriate PSD shapes. The X-ray PSD was modelled by a bending power-law with low frequency slope -1.1, high frequency slope -2.0 and break frequency $2 \times 10^{-4}$ Hz parameters from \citet{vaughan11a}. The observed rms-flux relation (see \citealt{uttleymchardyvaughan05}) was added to the simulated X-ray light curves by computing the exponential function of each point (\citealt{uttleymchardyvaughan05}; \citealt{vaughanuttley08}). The UV PSD was modelled with an unbroken power law with slope -2.1 (see section 3). Observational noise was added to each simulated UV and X-ray light curve by drawing a Poisson random deviate with mean equal to the mean count per bin in the real light curves. We then took 15 segments corresponding to the times and length of each real observation from the 50 day simulated light curves. The CCF was computed for each segment before averaging. This results in $10^{4}$ simulations of the averaged CCFs from which we extracted the confidence intervals at each lag. Confidence intervals on the cross-spectrum lag and coherence were computed using the same approach.

Light curves were simulated to put confidence intervals on the DCF using the above procedure, except the generated light curves were 20 times longer than before (1000 days). The time resolution was 1500 s for \textit{XMM} and 500 s for \textit{Swift} and 500 s for the combined datasets. A 50 day segment was then selected at random from this light curve and points then sampled from this coinciding with times and lengths of the real \textit{XMM} and \textit{Swift} observations.

In order to set an upper limit on coherence we simulated light curves with added variance i.e. some fraction \textit{A} of the simulated X-rays was added to the simulated UV. The variances were normalised before the two light curves were added. We simulated $10^{4}$ light curves in each band for a range added variance and recorded the mean coherence value for the lowest 5 Fourier frequencies. The bias in coherence (\citealt{bendatpiersol86}, section 9.2.3) is given by $B[\gamma^{2}] = 1/n(1 - \gamma^{2})^{2}$ where $\gamma^{2}$ is the coherence and \textit{n} is the number of segments going into the cross-spectrum (15 in our simulations). When the coherence is low the bias dominates and acts to shift up the observed coherence value and must be subtracted from the computed value. The distribution in coherence values for each \textit{A} were then compared to our observed coherence value $\gamma^{2}_{obs}$. The mean coherence value for the distribution of coherences where 90$\%$ of the values fall above $\gamma^{2}_{obs}$ is taken as the upper limit.

\end{document}